\documentclass{webofc}

\usepackage[varg]{txfonts}   
\begin{document}
\title{The Astrolabe Project: Identifying and Curating Astronomical ‘Dark Data’ through 
Development of Cyberinfrastructure Resources}
%
%
{}

\author{\firstname{Gretchen} \lastname{Stahlman}\inst{1}\fnsep\thanks{\email{gstahlman@email.arizona.edu}} \and
        \firstname{P. Bryan} \lastname{Heidorn}\inst{1}\fnsep\thanks{\email{heidorn@email.arizona.edu}} \and
        \firstname{Julie} \lastname{Steffen}\inst{3}\fnsep\thanks{\email{julie.steffen@aas.org}}
}

\institute{University of Arizona School of Information, 1103 E. 2nd St, Tucson, AZ 
\and
           American Astronomical Society, 667 K Street NW, Suite 800,
Washington, DC 
          }

\abstract{%
  As research datasets and analyses grow in complexity, data that could be valuable to other researchers and to support the integrity of published work remain uncurated across disciplines. These data are especially concentrated in the “Long Tail” of funded research, where curation resources and related expertise are often inaccessible. In the domain of astronomy, it is undisputed that uncurated "dark data" exist, but the scope of the problem remains uncertain. The “Astrolabe” Project is a collaboration between University of Arizona researchers, the CyVerse cyberinfrastructure environment, and American Astronomical Society, with a mission to identify and ingest previously-uncurated astronomical data, and to provide a robust computational environment for analysis and sharing of data, as well as services for authors wishing to deposit data associated with publications. Following expert feedback obtained through two workshops held in 2015 and 2016, Astrolabe is funded in part by National Science Foundation. The system is being actively developed within CyVerse, and Astrolabe collaborators are soliciting heterogeneous datasets and potential users for the prototype system. Astrolabe team members are currently working to characterize the properties of uncurated astronomical data, and to develop automated methods for locating potentially-useful  data to be targeted for ingest into Astrolabe, while cultivating a user community for the new data management system.
}
\maketitle
\section{Introduction}
\label{intro}
As a partnership between the American Astronomical Society and the University of Arizona School of Information, Astrolabe was formed in 2015 to address a demonstrated need for additional curation resources in astronomy. Based at the University of Arizona (UA), Astrolabe has thus far been funded by the University’s Research, Discovery and Innovation (RDI) division and the National Science Foundation (Advanced Cyberinfrastructure Directorate) to leverage existing cyberinfrastructure to create a new repository and computational environment for astronomical data that are not adequately curated in mission repositories or elsewhere. Additional collaborators at the UA include the Department of Astronomy and Steward Observatory, the University Libraries, and CyVerse (formerly known as the National Science Funded cyberinfrastructure project, the iPlant Collaborative). The organizational structure of this collaboration is further strengthened by the University of Arizona’s proximity to intellectual and technical resources, and its geographic centrality as a major hub for astronomy and information science research. The core mission of Astrolabe is to collect, preserve, and make accessible astronomical data that are not already successfully managed. The Astrolabe team is also developing tools for analysis and to encourage data sharing, as well as curation resources for authors of data and associated publications. Overall, the Astrolabe Project aims to help researchers expose data at key stages of the research lifecycle, such as the point of submitting a publication, for example, or when collecting new data for an existing project.
\section{Background}
\label{sec-1}
Popular data lifecycle models highlight the complexity of research across disciplines, particularly considering the deliberate re-use of scientific findings for subsequent research through established channels of review and dissemination as a key aspect of data lifecycles and an important priority for many disciplines and for scholarly publishing \cite{Ref1}-\cite{Ref4}. Additionally, with a tremendous amount of data being created, analyzed, and transformed in astronomy and contemporary science in general, the need for innovative methods of publishing and curating these data is well-documented, including both the technical and social aspects of data management challenges \cite{Ref5}-\cite{Ref10}. Pennock (2007) defines curation as “the active management and appraisal of digital information over its entire life cycle” \cite{Ref1}. While curation tends to bridge disciplinary boundaries, research has shown that curation also requires insightful knowledge of the data themselves - and of the scientific communities, technologies, analyses and activities that produce them - to facilitate best practices and provide researchers with appropriate scholarly and cyber-infrastructure. Furthermore, publication of both research and data is an important component of the data curation lifecycle, and it is critical to develop resources that support the needs of the community and the advancement of the discipline.
\subsection{“Dark Data” in the Long Tail of Astronomy}
\label{sec-2}
Much of astronomy’s data are curated in archives designed by and for large missions. However, much other data are known to be uncurated. Heidorn (2008) hypothesizes that large projects have well-planned data stores, while large amounts of data remain uncurated from the more numerous, small projects: “Like dark matter, this dark data on the basis of volume may be more important than that which can be easily seen” \cite{Ref11}. Heidorn’s exploration of research grants funded by National Science Foundation (NSF) across disciplines shows that the largest 20 percent of projects funded in 2007 received more than 50 percent of total funds awarded in that year. Heidorn suggests that this top 20 percent has well-curated data, primarily due to access to curation resources and budgeting for data management, and that the uncurated dark data distributed throughout the Long Tail - small projects of this distribution, which represents a major potential for accumulated knowledge if brought to light. Similar distribution of funding and presumably data curation resources are found in other fields such as ecology \cite{Ref12}.

A similar analysis of active awards in the Astronomy and Astrophysics Division of NSF in 2016 illustrates that while the characteristics of astronomy’s Long Tail are unique reflecting the extremely high funding given to a small number of large missions, a diversity of projects and levels of funding nonetheless exist that may require data curation services, to aggregate knowledge currently hidden from other researchers, and to support published research for the benefit of the astronomical community. Essentially, 20 percent of NSF-funded research projects in astronomy receive 90 percent of funding.
\newline

\begin{table}
  \centering
  \includegraphics[width=13cm,clip]{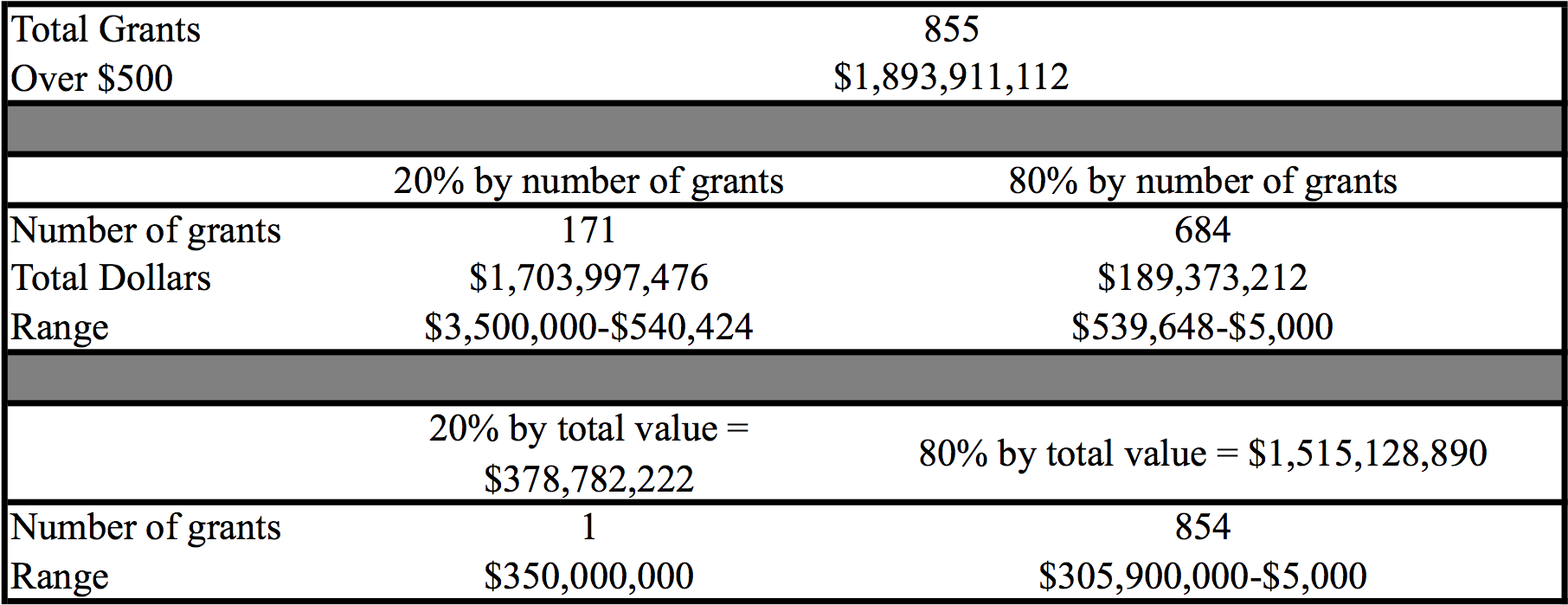}
  \caption{Grant Size Distribution of NSF Astronomy and Astrophysics Active Awards, running as of July 2016 }
   
\end{table}%
Astrolabe is intended to provide a repository and computational environment for smaller datasets and projects with less funding, and to make easy-to-use interfaces for uploading and interacting with data, and eventually integrating into the publication process.

Curating data in the Long Tail of any discipline is a challenge due to heterogeneity of data and a tendency towards lack of standard metadata.  Astronomical data types and formats correspond to particular research projects and observing techniques that vary widely across sub-disciplines and facilities. It is also often necessary to collect many observations over time, and astronomy faces growing challenges associated with generating, publishing, and utilizing very large and numerous datasets. Reprocessed or derived datasets and data collected by smaller instruments frequently remain uncurated. While the majority of data in astronomy conform to the Flexible Image Transport System (FITS) file format, with headers including coordinate information and details about instrumentation, many possible descriptive headers can exist, reintroducing heterogeneity. However, due to relatively successful adoption of FITS and other standards (including standards for interoperability developed by the International Virtual Observatory Alliance - IVOA - across the discipline and its global communities) astronomy is known for best practices in data management, and for widely supporting open access to scientific research.

\subsection{Astrolabe System Design}
\label{sec-3}
Astrolabe aims to support analysis and re-use of data in addition to collection and preservation services. Astrolabe is being constructed in partnership with CyVerse cyberinfrastructure, a platform for data sharing and analysis that was originally designed to support research in life sciences, but has expanded to other disciplines. CyVerse provides longevity to support data management planning, easy access to supercomputing resources and community-developed research tools, along with data federation through iRODS technology for portability and discoverability, and minting of permanent identifiers for static published data \cite{Ref13}. While anyone can obtain a CyVerse account and upload data, Astrolabe functions as a liaison between CyVerse and the astronomy community, facilitating development of functionality to make the system useful for astronomy data, and building on a foundation of tools and storage that include:

\begin{itemize}
\item{The “Discovery Environment” for convenient data and application sharing;}

\item{“Atmosphere” to launch virtual instances for analysis using High Performance Computing (HPC) resources; and} 

\item{A “Data Store” with ample free data storage (100GB per user).}
\end{itemize}

A substantial user community has formed within CyVerse, with more than 31,000 active users and nearly 3,000 Terabytes of user data, and projects representing a variety of disciplines across Natural and Life Science domains are now based in CyVerse technology \cite{Ref14}. 

Stakeholder-collaborators in the Astrolabe project are united by mutual need and interests and provide distinct assets to the organizational design of the project. This strategic collaboration includes: a School of Information (University of Arizona iSchool), representatives of a scientific discipline (astronomy), an academic library system (University of Arizona Libraries), a scholarly society (American Astronomical Society), and a cyberinfrastructure platform for cloud-based data storage and computation (CyVerse). As the publisher of four major journals in astronomy and astrophysics, with a global author community and with thousands of members from the United States astronomical community, American Astronomical Society stewards and supports other key elements of a broader vision to bring literature and data together and to progressively facilitate research and astronomy outreach, representing a key strength of the project. An advisory board has been established and convenes periodically to provide external guidance and to assist with establishing relationships with potential Astrolabe users and additional stakeholders. 

\section{Preliminary Research: Astrolabe Workshops}
\label{sec-4}
To begin planning and development for Astrolabe, two workshops were held in 2015 and 2016 to obtain expert feedback and to develop functional requirements for the system. Participants in the “Astrolabe” astronomy cyberinfrastructure workshops were primarily from the University of Arizona, National Optical Astronomy Observatory (NOAO), the American Astronomical Society (AAS) and CyVerse staff. Data collected from both workshops include: field notes taken by workshop organizers using a collaborative note-taking platform; preliminary surveys administered online using Qualtrics survey software; audio recordings of presentations, group discussions and breakout sessions; and transcription of key dialogue. This material was analyzed qualitatively by coding for key themes and sub-themes. 

\subsection{2015 Workshop Description and Outcomes}
\label{sec-5}
This workshop was held in Tucson in July 2015 with 32 participants and organizers, including astronomy and data science experts. Final recommendations of participants included:

\begin{itemize}
\item{Identify a mission statement and clear science use cases;}

\item{Take advantage of CyVerse cyberinfrastructure and the longevity of University of Arizona as a home for this new repository;}

\item{Obtain community buy-in and manage expectations for the system;}

\item{Focus on “low-hanging fruit” such as data not curated elsewhere and data behind figures in journals; and}

\item{Develop a follow-on workshop for additional feedback.}
\end{itemize}

\subsection{2016 Workshop Description and Outcomes}
\label{sec-6}
The 2016 “Astrolabe” workshop included 22 local and nonlocal participants. The group was comprised of research astronomers, University of Arizona and CyVerse representatives, and other expertise related to astronomy data and publishing, along with five workshop organizers. The following four topic areas were recommended for the immediate attention of Astrolabe:

\begin{itemize}
\item{Physical format of dark data (i.e. historical data stored on tapes);}

\item{Author websites archiving data (not typically long-lived);}

\item{LSST time domain and serendipitous data cases (such as follow-up to LSST observations and discovery through historical data); and}

\item{Searching the literature for references to dark data (for indicative text, broken links, etcetera).}
\end{itemize}

\section{Current Status and Ongoing Development of Astrolabe}
\label{sec-7}
Current activities of the Astrolabe Project include: 
\begin{itemize}
\item{Soliciting pilot datasets to develop workflows, and cultivating a user community;}

\item{Searching for uncurated or “at-risk” data by mining the literature, and by contacting authors individually based on the Astrolabe team’s review of particular types of publications;}

\item{Working towards the objectives specified in a National Science Foundation grant (award number 1642446) to develop WorldWide Telescope software as a repository front-end and visualization tool;}

\item{Submitting additional funding proposals for system development, including to broadly create protocols for migrating data from obsolete media into Astrolabe;}

\item{Collaborating with CyVerse to develop and optimize interfaces, applications, and indexing;}

\item{Developing a metadata pipeline, including an application for automatic extraction of some metadata from FITS headers, and creating a standard metadata template following the IVOA ObsCore model \cite{Ref15};}

\item{Incorporating the Unified Astronomy Thesaurus (UAT) as a controlled vocabulary \cite{Ref16}; and}

\item{Designing a web portal as an interface to the CyVerse data store, to conveniently facilitate data deposition and reuse.}
\end{itemize}

\subsection{WorldWide Telescope Interface}
\label{sec-8}
Originally developed by Microsoft Research but currently managed by the American Astronomical Society, WorldWide Telescope (WWT) is an open-source tool for visualization of astronomical data. With WWT, Astrolabe will provide users with a convenient option for automatically visualizing compatible image data within the repository environment and making Astrolabe datasets visible and accessible to WWT users. This new visualization functionality and cloud-based processing of image, all-sky and catalog data for display in WWT clients will be implemented over the next three years.

\subsection{Searching for “Dark Data” in the Literature}
\label{sec-9}
Attempts to locate publications of interest for Astrolabe follow-up have implemented combinations of search terms in the Astrophysics Data System (ADS) to narrow down the literature for manageable human review and selective follow-up with authors who may possess uncurated data. A list of papers without existing data links in the ADS index has been generated using the following search parameters:
\begin{itemize}
\item{24 articles contain the terms "new survey" or "new catalog" with a URL;}

\item{82 articles contain the term "SDSS-III";}

\item{188 articles contain the term "online data";}

\item{1399 articles published between 2006 and 2016 contain a URL.}
\end{itemize}

This spreadsheet provides additional information about the identified articles, including ADS bibcode identifiers, associated grant information, and number of times read online. As Astrolabe initially aimed to seek data from highly-cited Arizona authors, a second spreadsheet of articles lists 1142 articles associated with Arizona first-authors published between 2005 and 2015. A third study has identified broken links in journals published by AAS (a likely indicator of hidden data associated with a publication). Efforts continue to locate relevant papers within and beyond the four astronomy journals published by AAS. Machine learning and/or statistical modeling to determine significant variables associated with inaccessible data could assist the Astrolabe team with honing searching strategies to more accurately locate papers of interest, simultaneously providing insight into data sharing behavior and trends over time.

\section{Conclusion}
\label{sec-10}
The Astrolabe project builds on existing technology as much as is practical to provide preservation and access to otherwise uncurated data. The Astrolabe team is working with the scientific community to identify both the most critical datasets as well as the functionality that is needed for a new easy to use data and computing environment for the astronomical community.

%


\begin{thebibliography}{}
%
%
\bibitem{Ref1}
M. Pennock, Library and Archives \textbf{January} (2007)

\bibitem{Ref2}
J.C. Wallis, A. Pepe, M.S. Mayernik, C.L. Borgman, iSchools Conference (2008)

\bibitem{Ref3}
J. Greenberg, Cataloging and Classification Quarterly \textbf{47}, (2009)

\bibitem{Ref4}
P. Darch, A. Sands, iConference, http://hdl.handle.net/2142/73655 (2015)

\bibitem{Ref5}
T. Hey, S. Tansley, K. Tolle (Eds.), \textit{The Fourth Paradigm: Data-Intensive Scientific Discovery} (Microsoft Research, 2009)

\bibitem{Ref6}
A. Accomazzi, R. Dave, Astronomical Data Analysis Software and Systems XX, ASP Conference Proceedings \textbf{442}, (2011)

\bibitem{Ref7}
C.L. Borgman, Journal of the American Society for Information Science and Technology \textbf{63}, (2012)

\bibitem{Ref8}
C.L. Borgman, \textit{Big Data, Little Data, No Data: Scholarship in the Networked World.} (MIT Press, Cambridge, MA, 2015)

\bibitem{Ref9}
E. Henneken, Bulletin of the Association for Information Science and Technology \textbf{41}, 40-43 (2015)

\bibitem{Ref10}
A. Goodman, J. Fay, A. Muench, A. Pepe, P. Udomprasert, C. Wong, https://arxiv.org/abs/1201.1285 (2012)

\bibitem{Ref11}
P.B. Heidorn, Library Trends \textbf{57}, 280-299 (2008)

\bibitem{Ref12}
W.C.Lenhardt, M. Conway, E. Scott, B.O. Blanton, A. Krishnamurthy, M. Hadzikadic, M. Vouk, A. Wilson, J2016 IEEE High Performance Extreme Computing Conference (HPEC) (2016)

\bibitem{Ref13}
S.E. Hampton, C.A. Strasser, J.J. Tewksbury, W.K. Gram, A.E. Budden, A.L. Batcheller, C. Duke, J.H. Porter, Frontiers in Ecology and the Environment \textbf{11}, 156-162 (2013)

\bibitem{Ref14}
CyVerse, http://www.cyverse.org/learning-center (2017)

\bibitem{Ref15}
M. Louys, D. Tody, P. Dowler, D. Durand, L. Michel, F. Bonnarel, A. Micol, http://ivoa.net/documents/ObsCore/index.html (2017)

\bibitem{Ref16}
A. Accomazzi, N. Gray, C. Erdmann, C. Biemesderfer, K. Frey, J. Soles, https://arxiv.org/abs/1403.6656 (2014)

\end{thebibliography}
\end{document}